\documentclass[3p,times]{elsarticle}%

\journal{JSTAT}

\usepackage{amssymb}
\usepackage{amsmath}
\usepackage{amsthm}
\usepackage{graphicx}% Include figure files
\usepackage{bm}% bold math
\usepackage{hyperref}% add hypertext capabilities
\usepackage{xcolor}
\usepackage{float}
\biboptions{compress}

\begin{document}
	\begin{frontmatter}
		%Collective behaviors of stock market at different magnitude
		\title{The q-dependent detrended cross-correlation analysis of stock market}
		\author[add1,add2]{Longfeng Zhao\corref{cor1}}
		\ead{zlfccnu@mails.ccnu.edu.cn}
		\ead{zlfccnu@bu.edu}
		\author[add1]{Wei Li\corref{cor2}}
		\ead{liw@mail.ccnu.edu.cn}
		\cortext[cor1]{Corresponding author}	
		\cortext[cor2]{Corresponding author}	
		\author[add2,add6]{Andrea Fenu}
		\author[add2,add3,add4,add5]{Boris Podobnik}
		\author[add2,add7]{Yougui Wang}
		\author[add2]{H. Eugene Stanley}
		\address[add1]{Complexity Science Center\& Institute of Particle Physics, Hua-Zhong (Central China) Normal University, Wuhan 430079, China}
		\address[add2]{Center for Polymer Studies and Department of Physics, Boston University, Boston, Massachusetts 02215, USA}
		\address[add3]{Faculty of Civil Engineering, University of Rijeka, Rijeka, HR 51000, Croatia}
		\address[add4]{Zagreb School of Economics and Management, Zagreb, HR 10000, Croatia}
		\address[add5]{Luxembourg School of Business, Grand-Duchy of Luxembourg, Luxembourg}
		\address[add6]{Department of economics and management, University of Cagliari}
		\address[add7]{School of Systems Science, Beijing Normal University, Beijing 100875, PR China}
		\begin{abstract}
			The properties of q-dependent cross-correlation matrices of stock market 
			have been analyzed by using the random matrix theory and complex network. 
			The correlation structures of the fluctuations at different magnitudes have 
			unique properties. The cross-correlations among small fluctuations are much stronger than those among large fluctuations. The large and small fluctuations are dominated by 
			different groups of stocks. We use complex network representation to study these q-dependent matrices and discover some 
			new identities. By utilizing those q-dependent 
			correlation-based networks, we are able to construct some portfolio 
			by those most independent stocks which consistently perform the 
			best. The optimal multifractal order for portfolio optimization is
			approximately $q=2$. These results have deepened our understanding about the collective behaviors of the complex financial system.
		\end{abstract}
		
	\end{frontmatter}
\section{Introduction}
 
 Analysis of cross-correlations between different financial assets has become extremely attractive \cite{Plerou1999,Laloux1999}
 since the researchers started to report  the violation of Efficient Market Hypothesis (EMH).
 At the very beginning, the cross-correlation analyses  have  relied  on such linear  tools as the
Pearson correlation, which requires stationary in  the data, but real-world financial data sets are rarely stationary.
To take into account the non-linearity and non-stationarity  in real-world data,  new methods based on detrendization  have been proposed, among which the most popular has been the detrended fluctuation analysis
(DFA)\cite{Peng1994}. Motivated by  the DFA that is  applied  for a single time series, its generalization 
 named  detrended cross-correlation fluctuation analysis (DCCA) has been proposed to quantify the long-range cross-correlations between a pair of non-stationary signals \cite{Podobnik2008}. DFA and DCCA are subsequently extended by their  multifractal  versions: MFDFA and
MFDCCA, respectively\cite{Kantelhardt2002,Zhou2008,Oswiecimka2014}. DFA, DCCA and their 
multifractal counterparts have  been applied cross a broad range of systems including biological,
financial to physical systems\cite{Podobnik2011a,Salat2016,Zhao2017}. Recently an 
analog to the Pearson coefficient, the detrended cross-correlation coefficient 
$\rho(s)$ was introduced in Ref\cite{Zebende2011}. 
This coefficient applied to non-stationary signals 
 quantifies the significance level of correlations among fluctuations of detrended non-stationary
signals at a given detrending scale $s$ \cite{Kristoufek2014}. More recently the DCCA coefficient 
$\rho(s)$ has been widely used to study the non-linear cross-correlation among financial time series\cite{Podobnik2011a,Wang2013,Wang2013a,Zebende2013,Sun2016}.
Despite the success of the $\rho(s)$ coefficient, it has some limitations when  cross-correlations are  quantified 
among the fluctuations at different magnitude. A more recent extension of the $\rho(s)$, the q-dependent 
detrended cross-correlation coefficient $\rho(q,s), q\in R$, is based on the q-dependent fluctuation
functions $F_q$ from MFDFA and MFDCCA\cite{Kantelhardt2002,Oswiecimka2014,Kwapien2015a}. Kwapien et al. recently indicated that this method could be applied to the analysis of empirical data from such natural complex systems as 
physical, biological, social and financial systems. Our focus here is on financial market.

Here we  apply   the q-dependent cross-correlation coefficient to quantify the 
cross-correlation among the return time series of 401 constituent stocks of the S\&P 500 index. For those return
time series, we  generate  
  the q-dependent cross-correlation matrices $\mathrm{C}(q,s)$. We calculate 
 the statistical properties of the matrices at different multifractal orders and  varying time scales. As when analyzing the Pearson 
cross-correlation matrix, we analyze the eigenvalue and eigenvector dynamics of the 
matrices and find that the cross-correlations of stock market fluctuations at 
different magnitudes exhibit a unique structure and dynamics. The large fluctuations 
are always dominated by
a few industry groups, but the small fluctuations exhibit a different behavior.
We then represent the cross-correlations matrices as complex networks and use the planar maximally filtered 
graph (PMFG) method\cite{Tumminello2005} to construct the correlation-based networks and to analyze their basic topological features. The PMFG networks for small
fluctuations are more heterogeneous than  those obtained  for large fluctuations.
Using a centrality 
metric, we classify  stocks as  central or  peripheral  according to their 
centrality ranking. Applying this to portfolio optimization we find that a portfolio of peripheral stocks has a consistently higher return than one of central and randomly selected stocks.

The paper is organized as follows. In Sec. \ref{sec2} we  introduce  the methodology used in this paper. In Sec. \ref{sec3} we present 
the data and main empirical results. In Sec. \ref{sec4}, an application to portfolio optimization has been given. The last section provides our conclusions.

\section{Methodology\label{sec2}}
\subsection{q-dependent cross-correlation analysis}
The q-dependent cross-correlation coefficient can be obtained from the following procedure:\\
$(i)$ We consider a pair of time series $x_{i}$ and $y_{i}$, $i=1\ldots l$. We integrated these time series and obtain two new time series
\begin{align}
	\chi^x(k)=\sum\limits_{i=1}^{k} x_i-\langle x\rangle , k=1\ldots l ,\\
	\chi^y(k)=\sum\limits_{i=1}^{k}y_i  - \langle y\rangle, k=1\ldots l .
\end{align}
 $(ii)$ We divide $\chi^x(k)$ and $\chi^y(k)$ into $2M_s=2\times \mathrm{int}(l/s)$ non-overlapping boxes of length $s$ from the
 beginning and end of two integrated time series. We then calculate the the local trends for each segment $v (v=0,1,\ldots ,2M_s -1)$ by a least-square fit and subtract it from $\chi^x(k)$ and $\chi^y(k)$ to detrend the integrated series. We then find the residual 
 signals $X,Y$ equal to the difference between the integrated signals and the $m\mathrm{th}$-order polynomials $P_{s,v}^{(m)}$ fitted
 to these signals:
 \begin{align}
 X_v(s,i)&=\sum\limits_{j=1}^{i}\chi^x(vs+j)-P_{X,s,v}^{(m)}(j),\\
  Y_v(s,i)&=\sum\limits_{j=1}^{i}\chi^y(vs+j)-P_{Y,s,v}^{(m)}(j).
 \end{align}
 The covariance and variance of $X$ and $Y$ in a box $v$ are defined:
 \begin{align}
 	f^{2}_{XY}(s,v)&=\frac{1}{s}\sum\limits_{i=1}^{s} X_v(s,i)Y_v(s,i),\\
 	f^{2}_{ZZ}(s,v)&=\frac{1}{s}\sum\limits_{i=1}^{s}Z_v^2(s,i),
 \end{align}
 where $Z$ represents either $X$ or $Y$.\\
 $(iii)$ We then defined the fluctuation functions of order $q$ and scale $s$
 \begin{align}
 	F_{XY}^{q}(s)&=\frac{1}{2M_s}\sum\limits_{v=0}^{2M_s -1}sgn[f_{XY}^2(s,v)]|f_{XY}^2(s,v)|^{q/2},\\
 	F_{ZZ}^{q}(s)&=\frac{1}{2M_s}\sum\limits_{v=0}^{2M_s-1}[f_{ZZ}^2(s,v)]^{q/2}.
 \end{align}
The q-dependent cross-correlation coefficient between $x_{i}$ and $y_{i}$
 is defined:
\begin{align}
	\rho(q,s)=\frac{F_{XY}^{q}(s)}{\sqrt{F_{XX}^{q}(s)F_{YY}^{q}(s)}},
\end{align}
When $q=2$ we restore the detrended cross-correlation coefficient of $\rho(s)$\cite{Zebende2011}.
The q-dependent cross-correlation coefficient is bounded in $[-1,1]$ when $q>=0$. The coefficient 
can have arbitrary value when $q<0$. Here we focus on the case when $q>0$. The 
exponent $q$ acts as a filter. When $q>2$ the boxes with large fluctuations 
contribute to $\rho(q,s)$ the most, but when $q<2$ the
boxes with relatively small values dominate the fluctuation function, thus contribute to $\rho(q,s)$ the most.

\subsection{Random matrix theory}
Having introduced the q-dependent cross-correlation coefficient, we now construct the
cross-correlation matrices $\mathrm{C}(q,s)$ at different multifractal orders $q$ and detrending scales $s$. If we 
assume the correlation matrices are random, the random matrix theory can be employed as a benchmark 
to quantify to what extent the properties of q-dependent cross-correlation matrices deviate from the prediction of
purely random matrix. The random matrix theory has been widely applied  to investigate the collective phenomena in financial markets\cite{Laloux1999,Plerou1999,Gopikrishnan2001,Rosenow2002,Plerou2002,Podobnik2010,Zhou2012,Livan2011,Fenn2011,Wang2011d,Singh2015}. A comprehensive review is provided in Ref. \cite{Bun2016}. 

We consider a random correlation matrix constructed from time series, e.g., a return time series $r_i, i=1,\ldots L$\\
\begin{align}
	\mathrm{C}= \frac{1}{L}RR^{T},
\end{align}
where $R$ is an $N\times L$ matrix containing $N$ return time series $r_i$ of length $L$ with zero mean and unit variance, that are
mutually uncorrelated. The probability distribution function of eigenvalues of a random matrix can be written analytically in the limit
$N,L \longrightarrow \infty$ with a fixed $Q=\frac{L}{N}>1$
\begin{align}
	P(\lambda)=\frac{Q}{2\pi}\frac{\sqrt{(\lambda_{+}-\lambda)(\lambda-\lambda_{-})}}{\lambda},\label{eigen}
\end{align}
where $\lambda_{-}$ and $\lambda_{+}$ are the minimum and maximum eigenvalues of $\mathrm{C}(q,s)$. $\lambda_{-}$ and $\lambda_{+}$ are given by\\
\begin{align}
	\lambda_{\pm}=1+\frac{1}{Q}\pm 2\sqrt{\frac{1}{Q}}.
\end{align}
Equation \ref{eigen} is exact for Gaussian-distributed matrix elements. If the eigenvalue
distributions deviate from the prediction of \ref{eigen}, that signalizes the existence of mutual correlation in the time series.

We decompose the q-dependent cross-correlation matrices with eigenvalues 
$\lambda_{k},k=1\ldots N$ and eigenvectors $\mathrm{U_{k}},k=1\ldots N$ which 
provide information about the collective behavior of the stock market. Here we use 
the inverse participate ratio to quantify the reciprocal of the number of 
eigenvector components that significantly contribute. The inverse participate 
ratio (IPR) is defined
\begin{align}
	I_{k}=\sum\limits_{l=1}^{N}[u_{k}^{l}]^{4}
\end{align}
Here $u_{k}^{l}$ is the $l\mathrm{th}$ component of the eigenvector $\mathrm{U}_{k}$ corresponding to eigenvalue $\lambda_{k}$. 
The meaning of $I_{k}$ can be illustrated by two limiting cases, (i) a vector with identical components $u_{k}^{l}=1/\sqrt{N}$ has
$I_{k}=1/N$, whereas (ii) a vector with one component $u_{k}^{l}=1$ and the remainder 
zero has $I_{k}=1$. We also define the participate ratio (PR) as $1/I_{k}$, which approximately equal to the significant contributors for eigenvalue $\lambda_{k}$. In random matrix theory, the expectation of IPR is
\begin{align}
	\langle I_k\rangle=N\int\limits_{-\infty}^{\infty}[u_{k}^{l}]^4\frac{1}{\sqrt{2\pi N}}\mathrm{exp}(-\frac{[u_{k}^{l}]^2}{2N})\mathrm{d}u_{k}^{l}=\frac{3}{N}.
\end{align}
%We study the short-range correlation by the nearest-neighbor and next to nearest-neighbor spacing distribution for unfolded eigenvalues. The long-range correlations among eigenvalues are investigated by
%using the number variance.
%The unfolded eigenvalues $\xi_{i}$ has uniform distribution. The nearest-neighbor spacing
%$d=(\xi_{i+1}-\xi_{i})$ for the GOE is:
%\begin{align}
%	P_{GOE}(d)=\frac{\pi d}{2}\mathrm{exp}(-\frac{\pi}{4}d^{2})
%\end{align}
%The next nearest-neighbor spacing for GSE is:
%\begin{align}
%	P_{GSE}(d)=\frac{2^{18}}{3^{6}\pi^{3}}d^{4}\mathrm{exp}(-\frac{64}{9\pi}d^2)
%\end{align}
\subsection{Planar maximally filtered graph}
As suggested in Ref.\cite{Kwapien2015a}, we use the complex network approach to analyze the q-dependent cross-correlation matrix.
We employ the the planar maximally filtered graph (PMFG) method \cite{Tumminello2005} to
construct networks based on correlation matrices $\mathrm{C}(q,s)$. The algorithm
is implemented as follows,\\
(i) Sort all of the $\rho_{ij}(q,s)$ in descending order to obtain an ordered list $l_{sort}$.\\
(ii) Add an edge between nodes $i$ and $j$ based on the order in $l_{sort}$ only when the graph remains planar after the edge is added.\\
(iii) A graph $G(q,s)$ is formed with $N_e=3(N-2)$
edges under the constraint of planarity.\\
\indent As described in Ref.\cite{Tumminello2005}, PMFGs not only keep the hierarchical
organization of the minimum spanning tree (MST) but also generate cliques. We calculate
the basic topological parameters such as clustering coefficient $C$, the shortest-path length
$L$ and the assortativity $A$. We also adopt a heterogeneity index $\gamma$ \cite{estrada2010quantifying} to measure the heterogeneity of PMFGs which is defined by
\begin{equation}
\gamma=\frac{N-2\sum\limits_{ij\in {\{e\}}}(k_{i}k_{j})^{-1/2}}{N-2\sqrt{N-1}},
\label{heterogenity}
\end{equation}
Here $k_i$ and $k_j$ are the degrees of nodes $i$ and $j$ connected by edge $\{e_{ij}\}$. 
%We have also investigate the community and sector structures for different detrended scale and fluctuation order. Some hidden structure has been uncovered.
\section{Data and Results\label{sec3}}
\subsection{Data description}
Our data sets include $N=401$ S\&P500 constituent stocks from 4 January 1999 to 31 December 2014 with 4025 trading records for each stock. We use the logarithm return defined as
\begin{equation}
r_{i}(t)=\mathrm{ln}p_{i}(t+1)-\mathrm{ln}p_{i}(t),
\end{equation}
\noindent
where $p_{i}(t)$ is the daily adjusted closure price of stock $i$ at time $t$. We then
use the previous method to compute the q-dependent cross-correlation coefficients between any pair of return time series $r_i(t)$ and $r_j(t)$ and obtain the $N\times N$ matrix $\mathrm{C}(q,s)$. The matrix entries of $\mathrm{C(q,s)}$ are the correlation coefficients $\rho_{ij}(q,s)$ between all pairs of
stocks. We set $q\in [0.2,5]$ with $\delta q=0.2$ and the detrending scale $s\in [30,1000]$ trading days with $\delta_s=40$.
We also perform the same calculation on the shuffled return time series and  the simulated random time series and use them as reference models.
\subsection{Cross-correlation matrix analysis}
With a series of cross-correlation matrices $\mathrm{C}(q,s)$ at different order $q$ and detrending scale $s$, we analyze the probability distribution of the cross-correlation values, i.e, the upper triangle entries of the correlation matrices.
\begin{figure}
	\includegraphics[width=\textwidth]{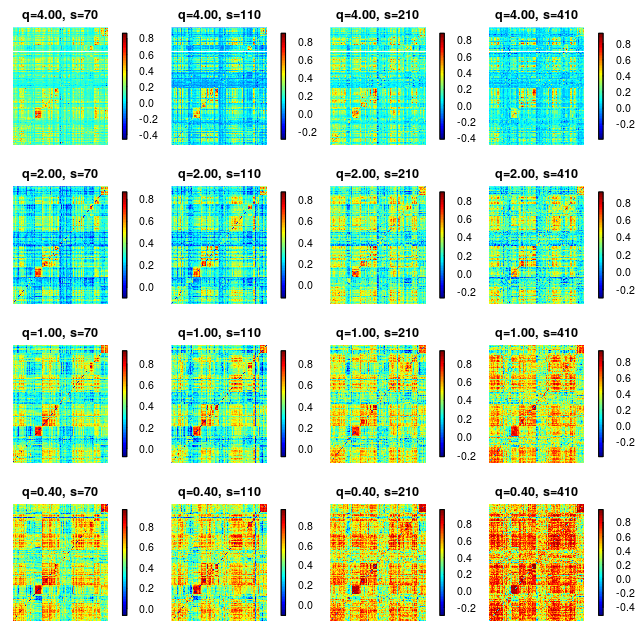}
	\caption{\label{mat}(Color on line) The cross-correlation matrices for different order $q$ and scale $s$. The diagonal elements have been set to zero. }
\end{figure}

First we show the plot of matrices for different multifractal order $q$
and detrending scale $s$ in Fig \ref{mat}, set the diagonal 
entries to zero for better visualization. The strength of the average correlation 
will increase slightly
as the scale $s$ increases, but will decrease as the multifractal order $q$ increases. We sort the rows and columns of the correlation matrices according to the official sector and subsector partitions of S\&P500.
Note the distinct sector and subsector structures in the correlation matrices.
When $q<2$ the sector structure is much more pronounced.
\begin{figure}
	\includegraphics[width=\textwidth]{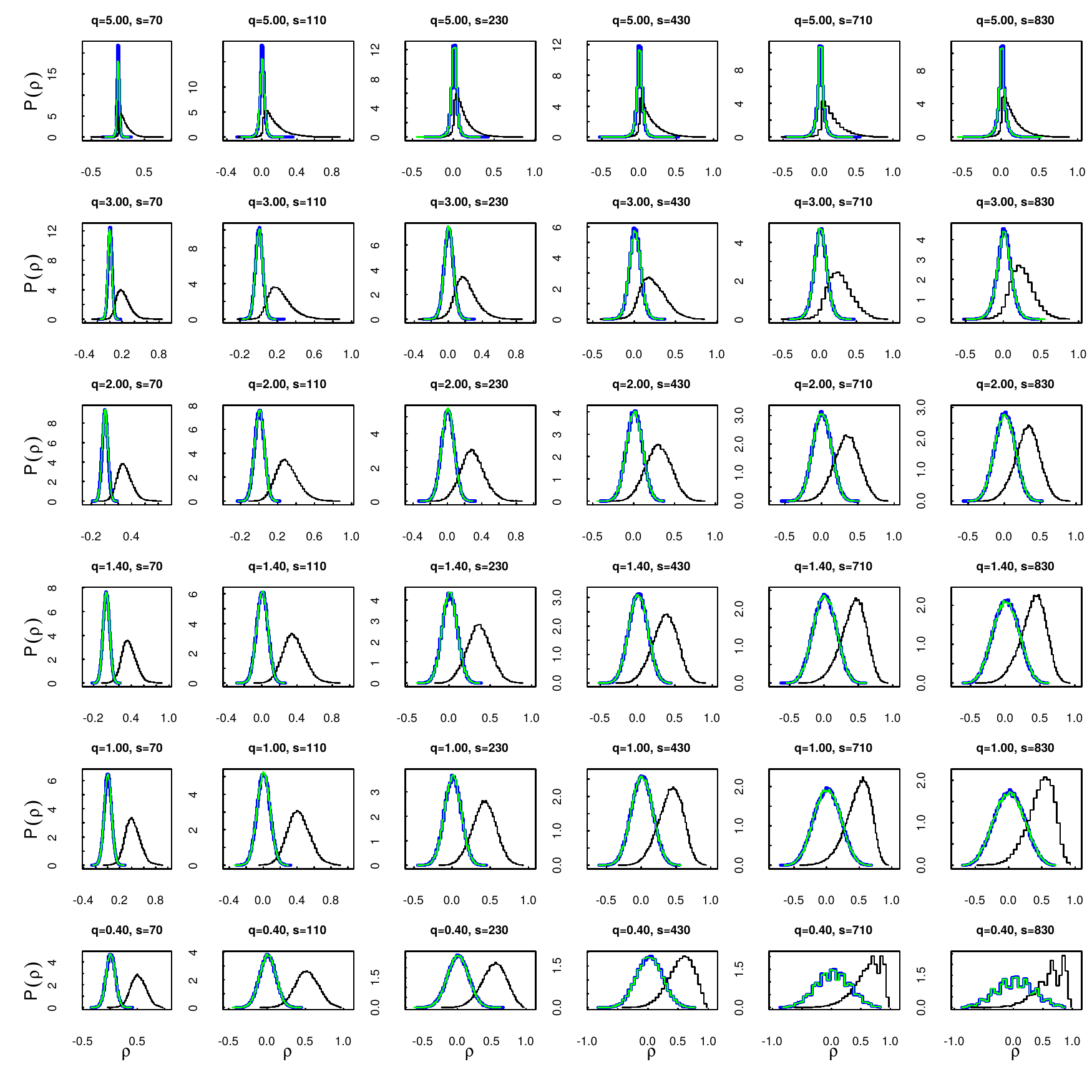}
	\caption{\label{dist}(Color on line) The distribution of the non-diagonal elements of the cross-correlation matrices. The black lines are the distribution of the q-dependent cross-correlation for the return time series. The blue lines and green lines are the same distribution for the shuffled
		return time series and simulated random time series, respectively.}
\end{figure}

Fig\ref{dist} shows the distribution of the matrices elements $\mathrm{P}(\rho)$ for six different values of $q$ and six different
values of scale $s$. We can observe that the distribution of the matrices become increasingly skewed to the left and the width 
of the distribution peaks as the multifractal order $q$ increases. The probability distribution of the q-dependent cross-correlation coefficient for the return time series deviate significantly from the shuffled distribution, and this may provide genuine information about the cross-correlation among different magnitude of fluctuations.
The shuffled and simulated distributions are coincide with each other.
Thus the different cross-correlation structure is the result of the non-linear
correlation among different magnitude of fluctuations.
In addition, when $q>2$ the distribution becomes relatively close to the shuffled case. We calculate the first four order moments of the correlation matrices to illustrate the variation in the cross-correlation distribution.

\begin{figure}
	\includegraphics[width=\textwidth]{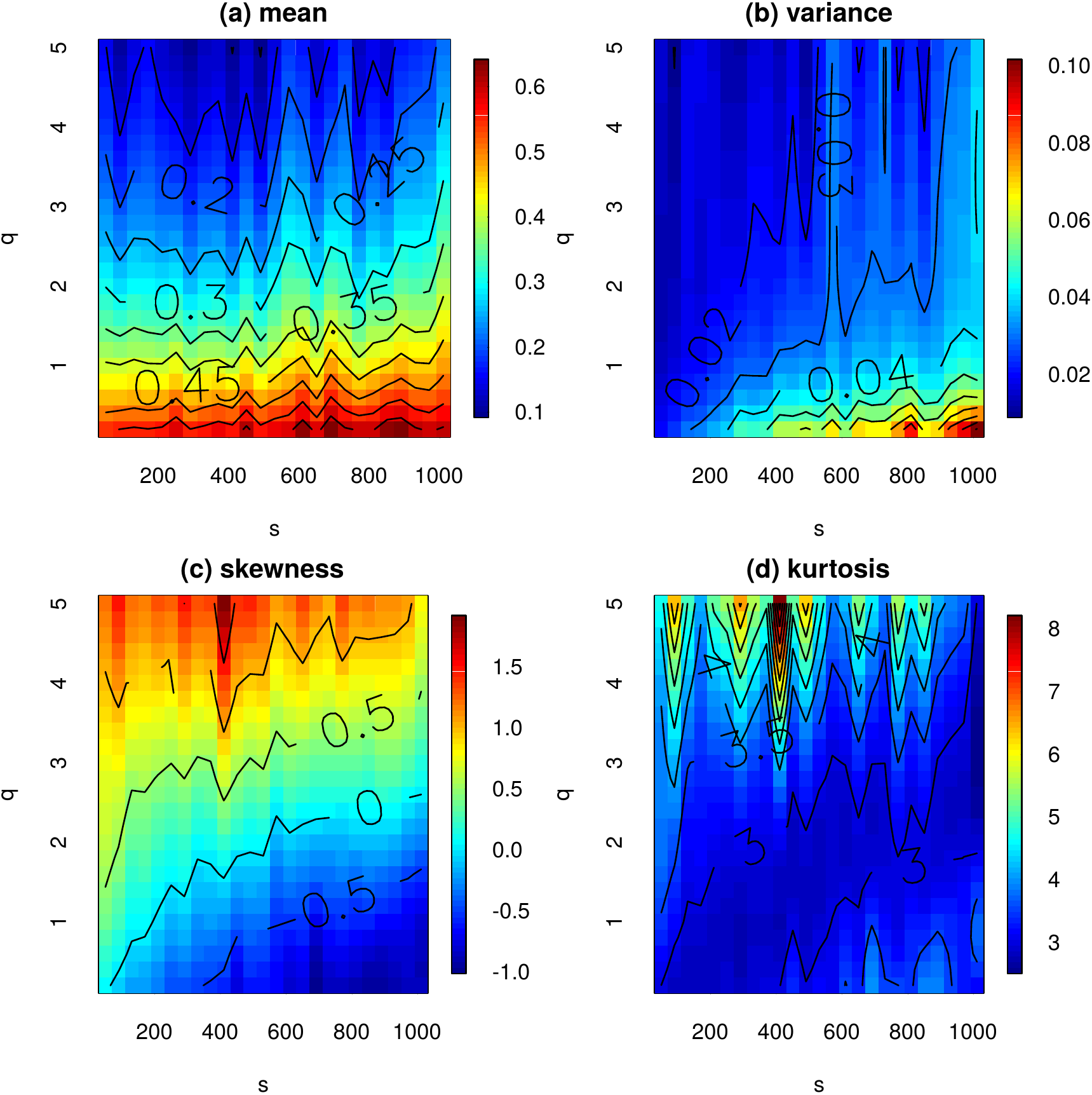}
\caption{\label{statsinfo} (Color on line) First four order moments: mean, variance, skewness, kurtosis of the correlation matrices at different multifractal order $q$ and detrending scale $s$.}
\end{figure}

Fig. \ref{statsinfo} shows the first four order moments of the correlation coefficient distribution at different multifractal orders $q$ and detrending scales $s$. 
The average cross-correlation 
decreases as the multifractal order increases, indicating that the cross-correlation between large fluctuations are relatively weak. From the variance, skewness and kurtosis we see an obvious transition in the distribution. The cross-correlation coefficients for large and small multifractal orders $q$ are largely 
different, which indicate disparate correlation structures among different magnitude of fluctuations.

\begin{figure}
	\includegraphics[width=\textwidth]{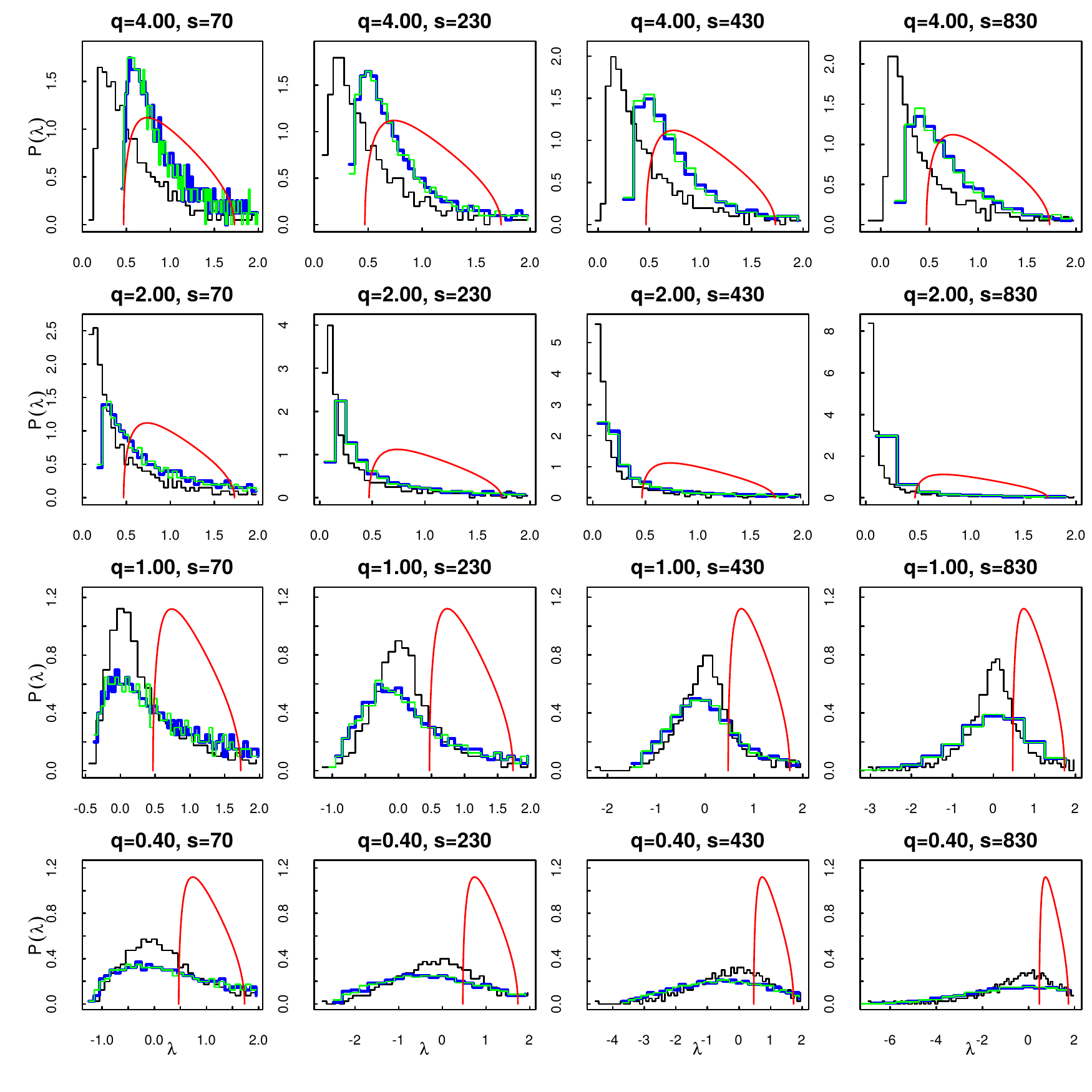}
	\caption{\label{EigenBulk}(Color on line) The eigenvalues distribution $\mathrm{P}(\lambda)$ of the cross-correlation matrices inside the bulk. We only show the distribution of those eigenvalues smaller than 2. The black, blue and green lines are the eigenvalue distributions of the q-dependent cross-correlation matrices for the original, shuffled and simulated time series ,respectively.}
\end{figure}

\begin{figure}
	\includegraphics[width=\textwidth]{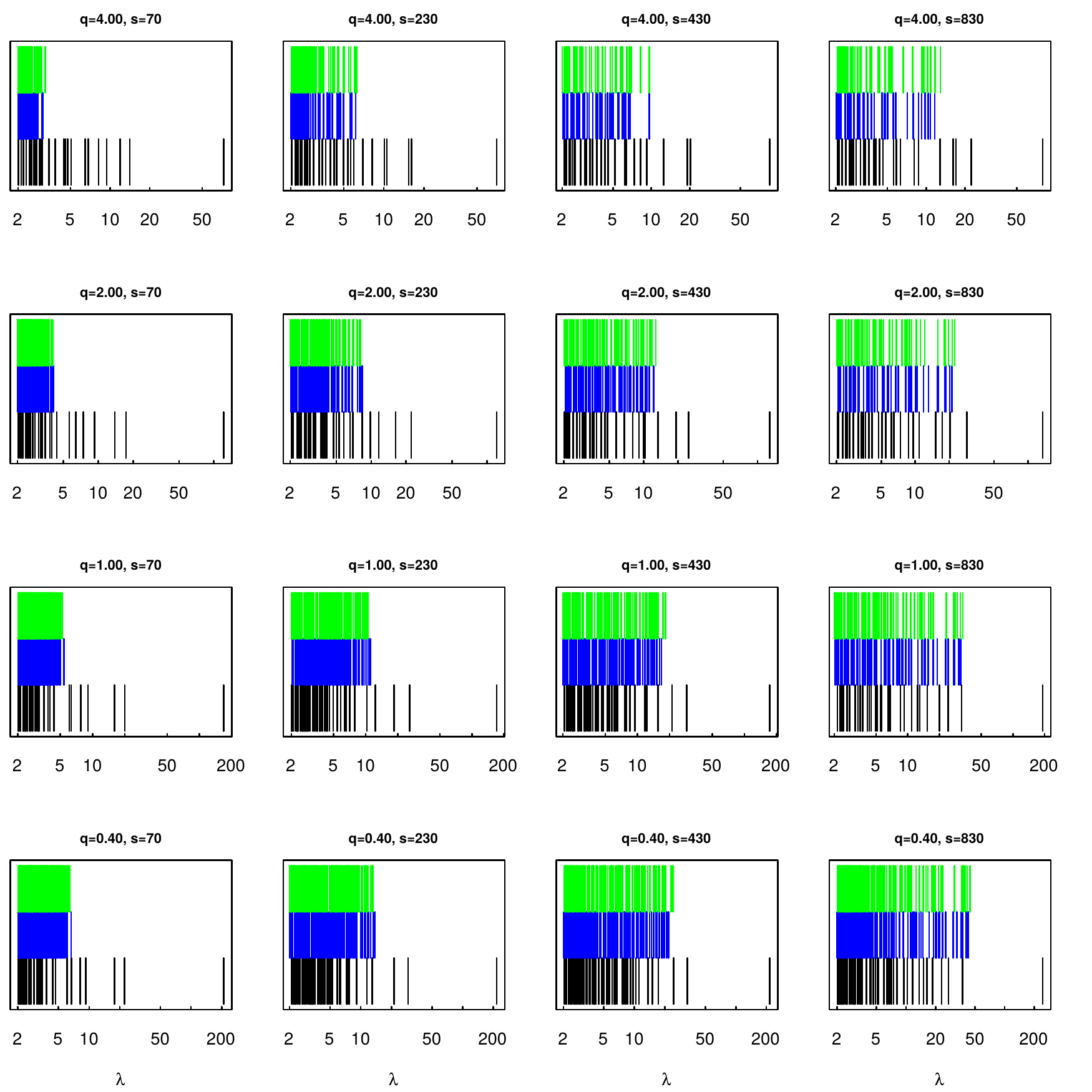}
	\caption{\label{EigenDeviate}(Color on line) The eigenvalues of the cross-correlation matrices deviating from the bulk ($\lambda >2$). The meaning of the color is the same as in Fig\ref{EigenBulk}.}
\end{figure}

To analyze the genuine information carried by the q-dependent cross-correlation 
matrices, we decompose the cross-correlation matrices and sort the eigenvalues 
$\lambda_{k},k=1\ldots 401$ in ascending order with their corresponding eigenvectors 
$\mathrm{U}_{k},k=1\ldots401$. Fig \ref{EigenBulk} and Fig \ref{EigenDeviate} show 
the distributions of the bulk eigenvalues and deviating eigenvalues, respectively. 
Fig. \ref{EigenBulk} only gives eigenvalues smaller than 2. The black and blue
lines are the eigenvalue distributions for the original q-dependent cross-correlation 
matrices and the shuffled scenario. The red lines are the 
eigenvalue distributions predicted by random matrix theory. 
We also simulate 401 time series using Gaussian distribution. The green lines are
the bulk eigenvalues from the q-dependent cross-correlation matrices calculated 
using the simulated Gaussian time series. We find that the bulk eigenvalue 
distribution of the shuffling time series and the simulated time series are 
approximately the same. This confirms 
that the deviation of the eigenvalue distribution is the result of non-linear 
cross-correlation.
 The lower and upper bounds of the eigenvalues predicted by RMT are $\lambda_{-}=0.47$ and 
 $\lambda_{+}=1.73$.
The distribution of the bulk eigenvalues for the original q-dependent cross-correlation differ from the random matrix
theory prediction. Note that when $q>2$ the bulk eigenvalue distribution for the original cross-correlation matrices
and the shuffled matrices approaches the random matrix prediction.
Fig. \ref{EigenDeviate} shows the deviating eigenvalues for the original cross-correlation matrices (black), the shuffled results (blue), and the simulated 
results (green). The behavior of those deviating eigenvalues differs as the values of 
$q$ and $s$ differ. Large $q$ values and small $s$ values tend to cause more large deviating eigenvalues.
Note that the deviating 
eigenvalues for large $q=4$ and small $s=70$ are especially clear.
In contrast, when $q=0.4$ and $s=830$ only the largest eigenvalue continues to deviate from
the shuffled and simulated eigenvalues. This indicates that the small fluctuations 
only have a very short characteristic time. The long term average effect of small fluctuations equals the noise level. Generally speaking, large multifractal order $q$ and small detrending scale $s$ makes the sector structures (deviating eigenvalues) and 
market mode (largest eigenvalue) separated from the noise level.

\begin{figure}[ht]
	\includegraphics[width=\textwidth]{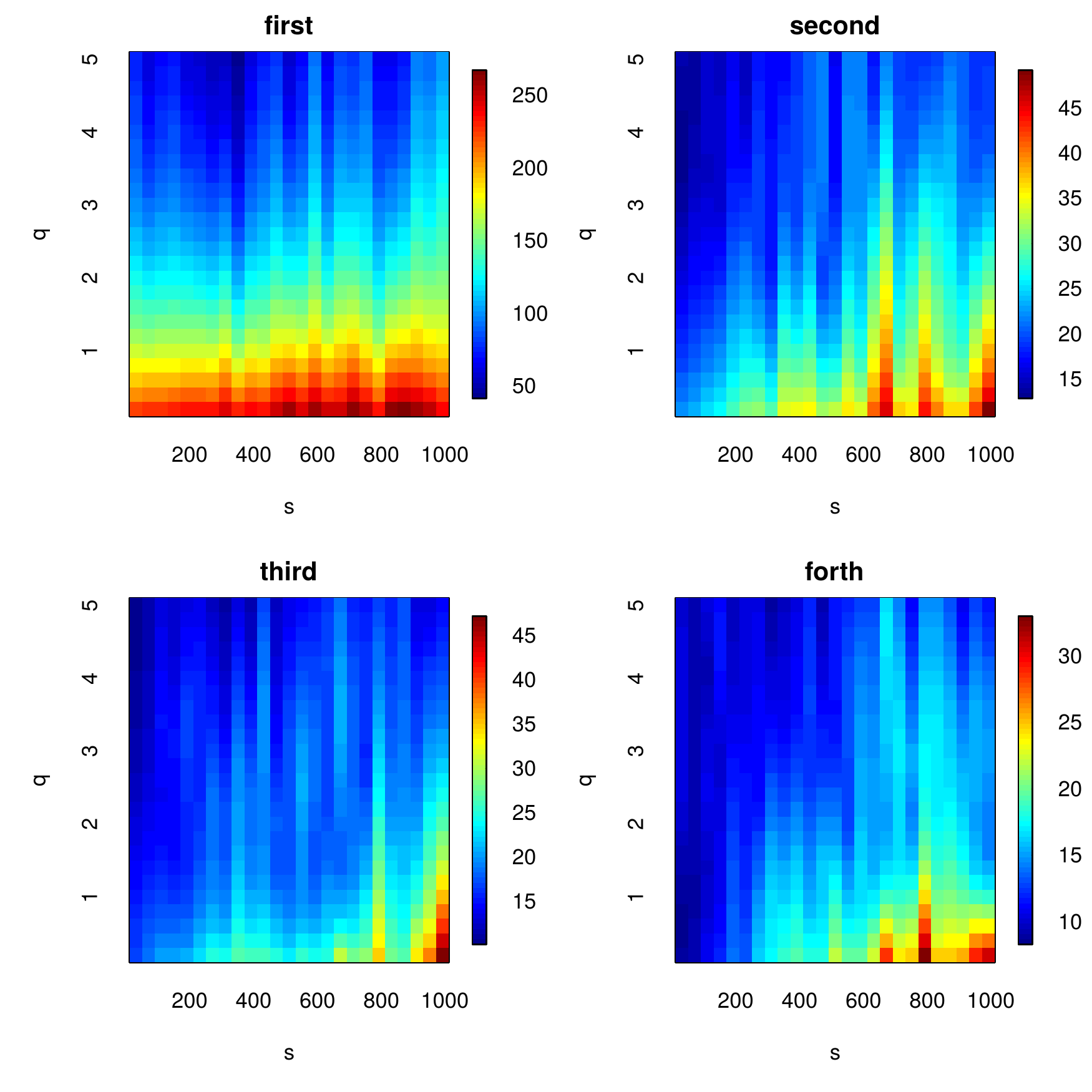}
	\caption{\label{1-4Eigen}(Color on line) The first four eigenvalues $\lambda_{K},k=1\ldots4$ as a function of multifractal order $q$ and detrending scale $s$.}
\end{figure}
The first four eigenvalues for different multifractal orders $q$ and detrending scales $s$ is shown in Fig \ref{1-4Eigen}. The largest eigenvalues for $q<2$ are approximately equal to
the order of the system size. The behavior of the largest eigenvalues is similar to 
the average cross-correlation in Fig. \ref{statsinfo}(a). This support the conclusion that 
the largest eigenvalue corresponds to the market mode described by numerous researches\cite{Gopikrishnan2001,Plerou1999} and it decreases when the value of $q$ increases. Thus the market mode at small $q$ is extremely stronger, which seems counterintuitive. We also observe that the first four eigenvalues increase
as detrending scale $s$ increases.

\begin{figure}[ht]
	\includegraphics[width=\textwidth]{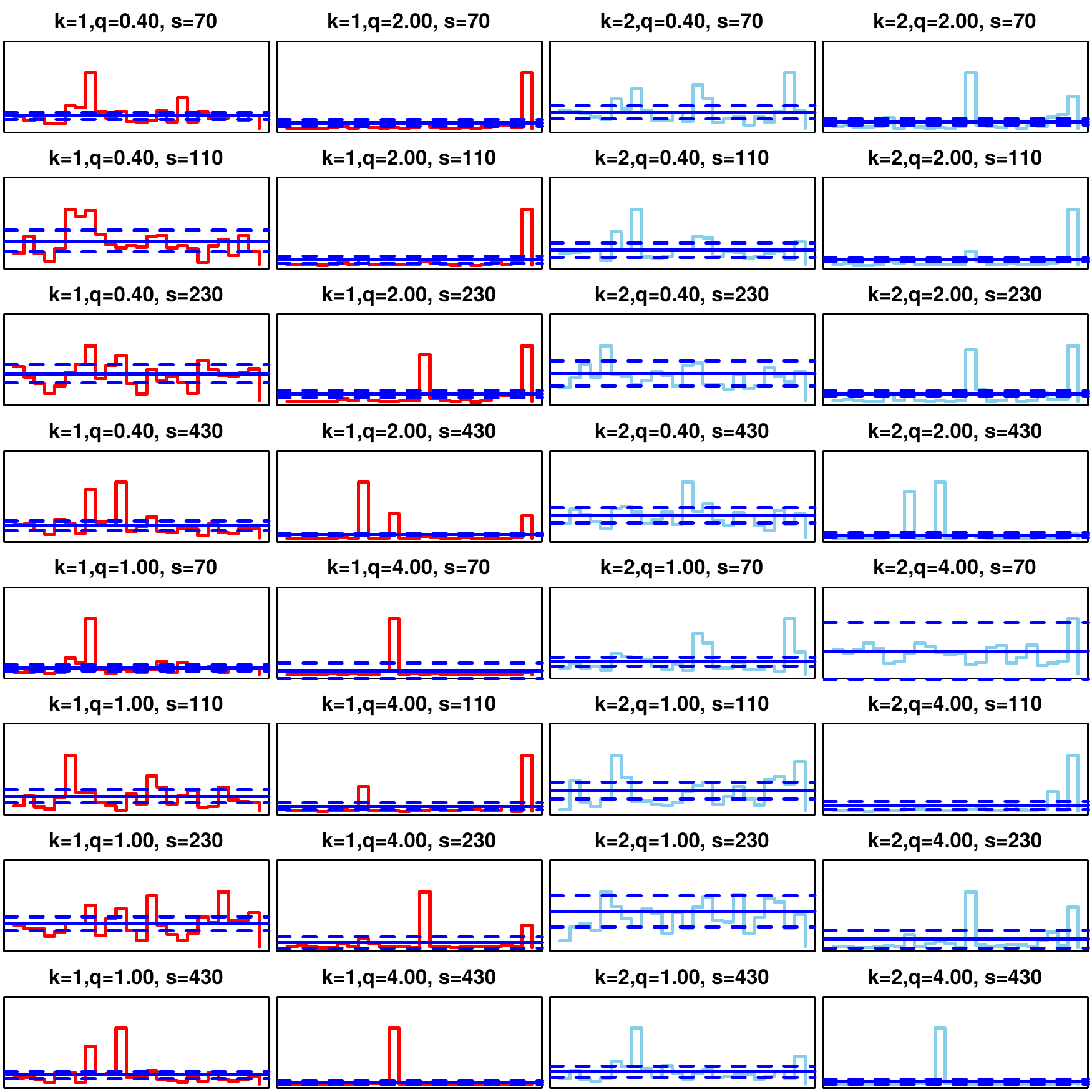}
	\caption{\label{eigenComp1-2} (Color on line) The contribution $X_{k}^{l},l=\ldots 24$ of every industry group to the smallest eigenvalue $\lambda_{k},k=1$ (red lines) and second smallest eigenvalue $\lambda_{k},k=2$ (sky blue lines) at different multifractal order $q$ and detrending scale $s$. The blue solid and dashed lines are the mean $X_{k}^{l}$ with one standard deviation for the shuffled correlation matrices.}
\end{figure}

\begin{figure}[ht]
	\includegraphics[width=\textwidth]{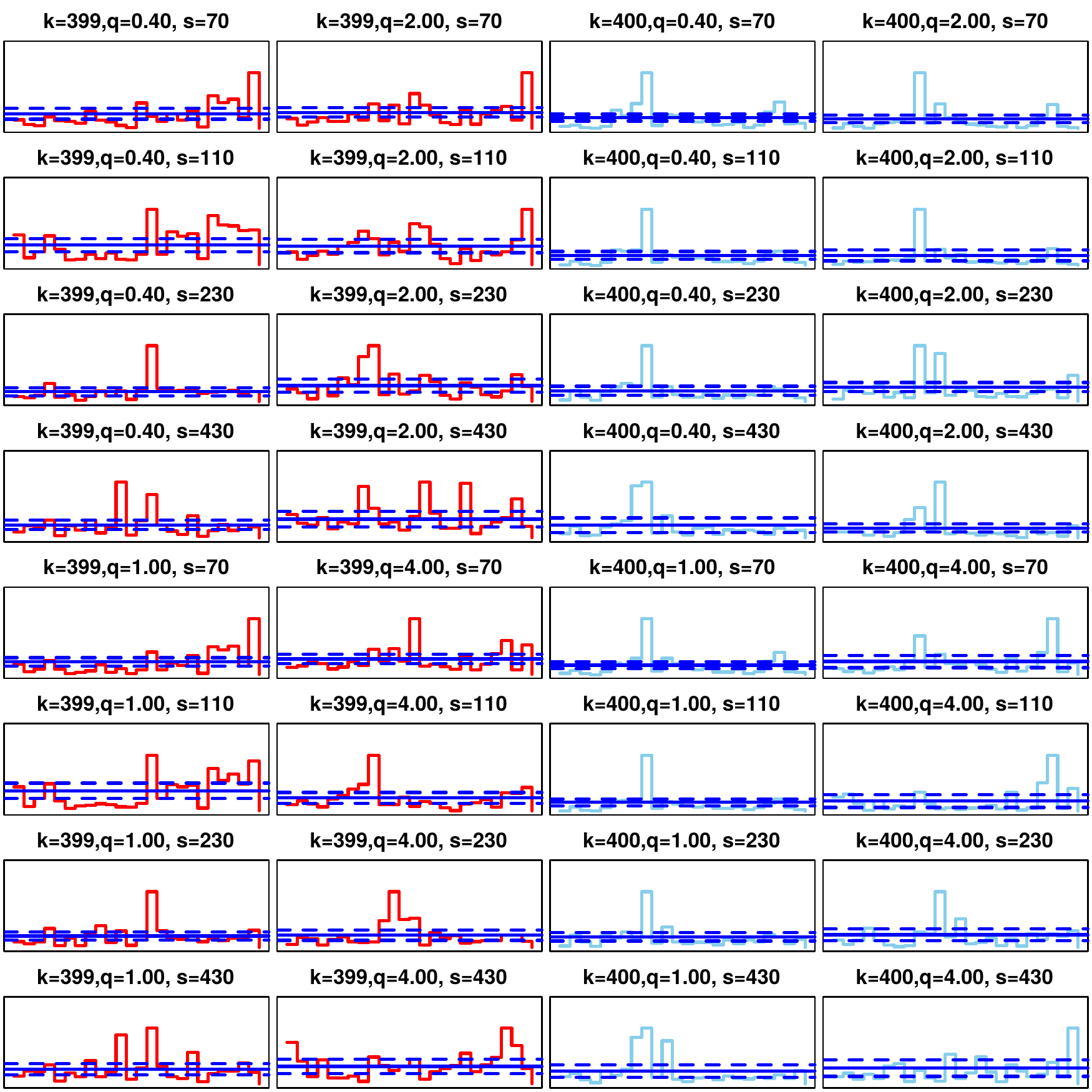}
	\caption{\label{eigenComp399-400} (Color on line) The contribution $X_{k}^{l},l=1\ldots 24$ of every industry group to the third largest eigenvalue $\lambda_{k},k=399$ (red lines) and second largest eigenvalue $\lambda_{k},k=400$ (sky blue lines) at different multifractal order $q$ and detrending scale $s$. The blue solid and dashed lines are the mean $X_{k}^{l}$ with one standard deviation for the shuffled correlation matrices.}
\end{figure}

It is believed that those eigenvalues deviated from the prediction of random matrix theory
contains some genuine information related to the sector or industry as described in Ref. \cite{Gopikrishnan2001,Plerou2002}. To 
uncover the hidden information carried by those deviating eigenvalues at different multifractal orders and detrending scales, we first partition the 401 stocks into industry groups labeled $l=1\ldots 24$ ($N_l$ stocks each) according 
to the industry group code of the stocks supplied by GICS. We then construct a projection matrix $P$, with elements $P_{li}=1/N_{l}$ if stock
$i$ belongs to industry group $l$ and $P_{li}=0$ otherwise. For each eigenvector $\mathrm{U}_{k}$, the contribution $X_{k}^{l}=\sum\limits_{i=1}^{N}P_{li}(u_{k}^{i})^2$ of each industry group
can be obtained.
Fig \ref{eigenComp1-2} shows the contribution of each industry group to the smallest and  second
smallest eigenvalues $\lambda_{k},k=1$, $\lambda_{k},k=2$. The red ($k=1$) and sky blue ($k=2$) lines are the contribution value after 
the influence of the largest eigenvalue for $\lambda_{401}$ is removed. The blue lines are the average contribution value $X_{k}^{l}$ for the correlation matrices calculated using the shuffled time series. This reference model tells us how much the $X_{k}^{l}$ deviate from the noise 
level. There are 24 major industry groups for the 401 stocks: \textbf{Media, Retailing, 
Consumer Durables\&Apparel, Automobiles\&Components, Consumer Services, Food\&Beverage\&Tobacco, Food\&Staples Retailing, Household\&Personal Products, Energy, Diversified Financials, Banks, Insurance, Real Estate, Pharmaceuticals, Biotechnology\&Life Sciences, Health Care Equipment \& Services, Capital Goods, Transportation, Software\&Services, Commercial\&Professional Services, Materials, Technology Hardware\&Equipment, Semiconductors\&Semiconductor Equipment, Telecommunication Services, Utilities}, from left to right.
 It is shown that for the smallest and second smallest eigenvalues, $\lambda_{1}$ and $\lambda_2$, the contribution come
from a few industry groups and the $X_{k}^{l}$ for these 
industry groups are much stronger than the noise level. The
industry contribution of the large eigenvalues $\lambda_{k},k=399, 400$ are presented in  Fig. \ref{eigenComp399-400}. The contribution to large eigenvalues also come from 
a few industry groups and is much stronger than the noise level. For $\lambda_{399}$,
there are multiple industry contribute significantly with a mixture pattern. The main contribution come from Diversified Financials, Banks, Real Estate and Utilities. But the
contribution to $\lambda_{400}$ always comes from Energy and Utilities.

\begin{figure}[ht]
	\includegraphics[width=\textwidth]{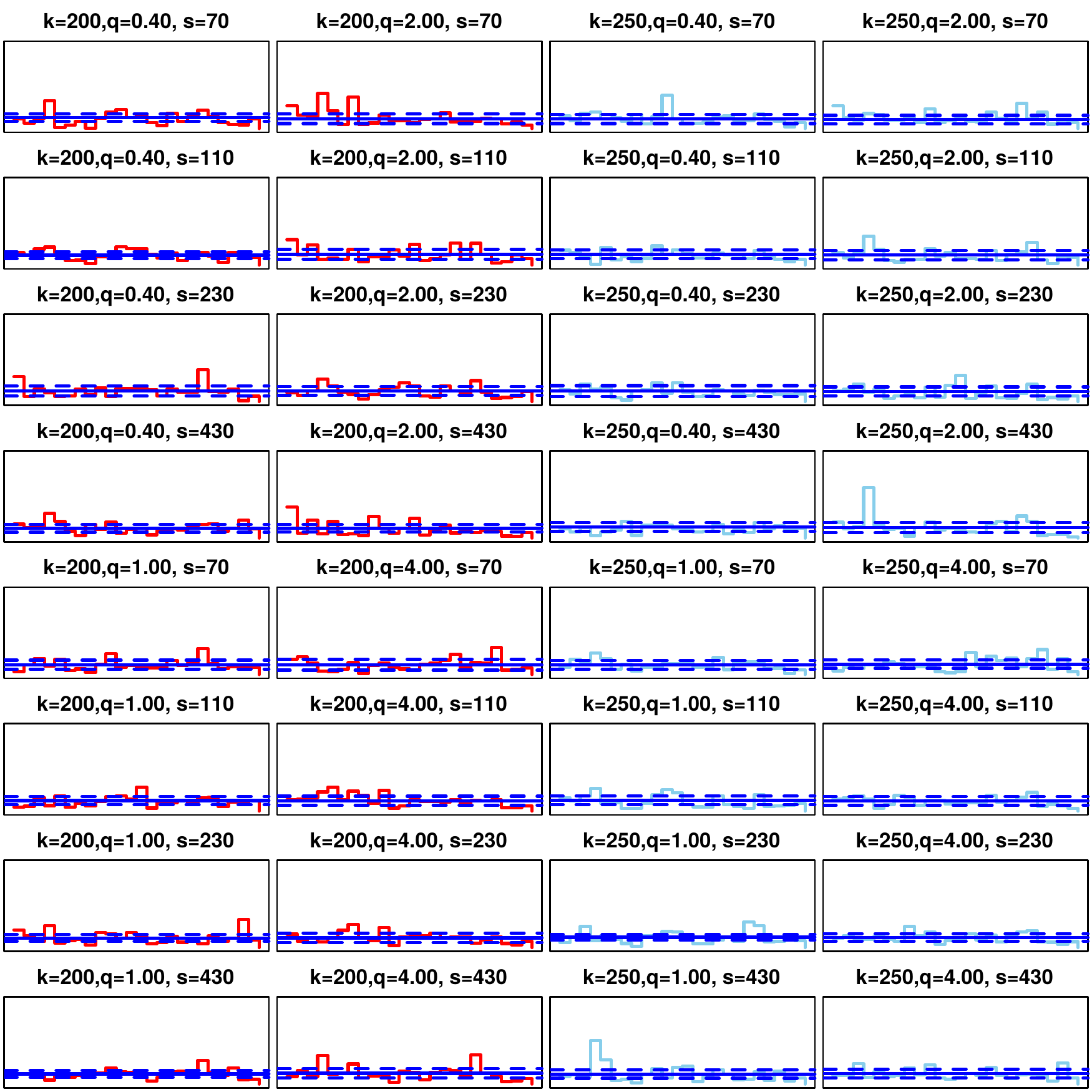}
	\caption{\label{eigenComp200-250}(Color on line) The contribution $X_{k}^{l},l=\ldots 24$ of each industry group to those eigenvalues fall deep inside the bulk $\lambda_{k},k=200$ (red lines) and $\lambda_{k},k=250$ (sky blue lines) at different order $q$ and detrending scale $s$. The blue solid and dashed lines are the mean $X_{k}^{l}$ with one standard deviation for the shuffled correlation matrices.}
\end{figure}
As shown in Fig\ref{EigenBulk}, there are many small eigenvalues within the 
prediction of random matrix theory. Fig. \ref{eigenComp200-250}
shows the contribution of each industry group to the eigenvalues $\lambda_{k},k=200,250$ deep inside the
eigenvalue bulk region. As expected, in this region the eigenvalues exhibit no significant pattern. The contribution level $X_{k}^{l}$ of each industry group is the same as in the shuffled time series. 
For both $\lambda_{k},k=200$ and $\lambda_{k},k=250$, there are no clearly contributing industry groups.

 \begin{figure}[ht]
 	\includegraphics[width=\textwidth]{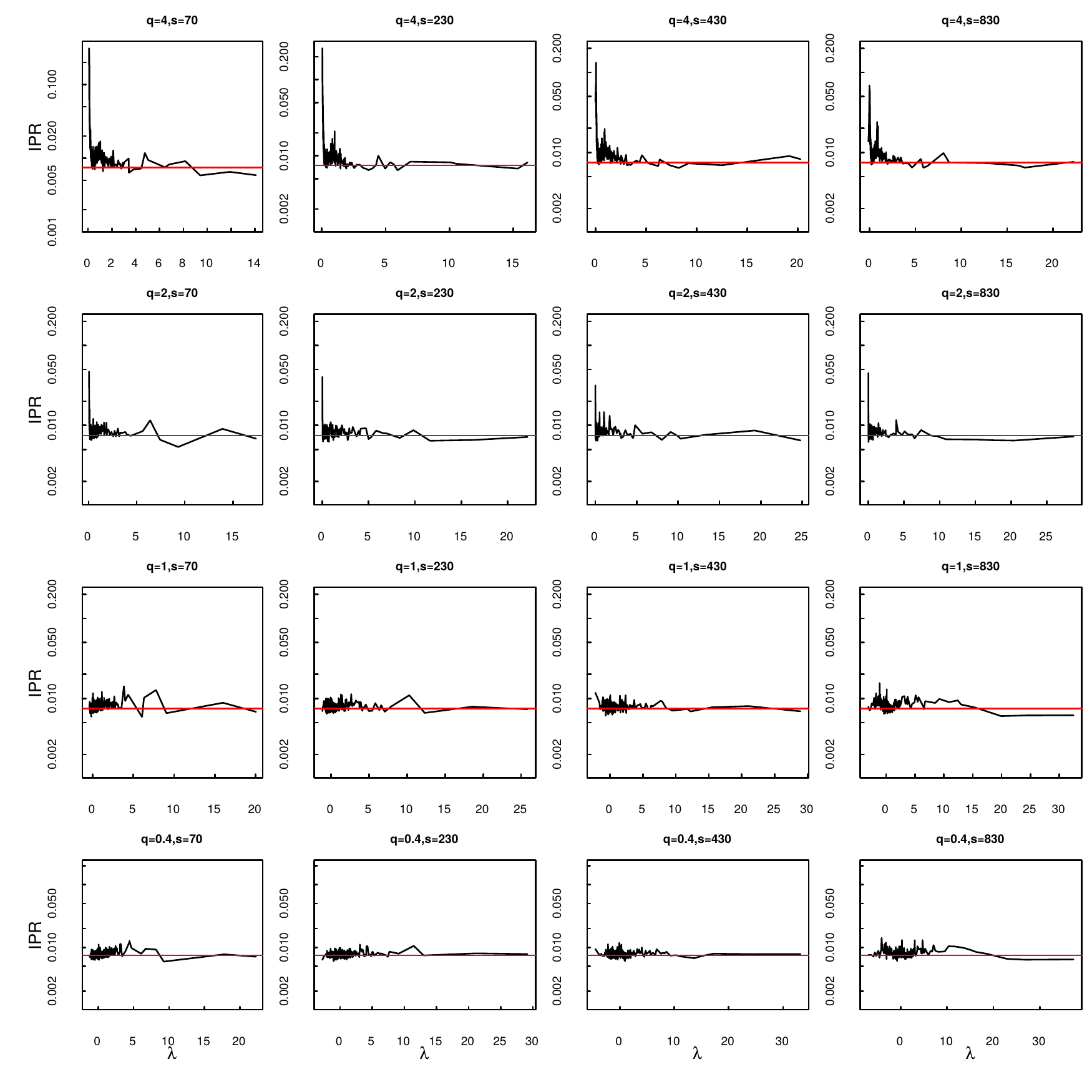}
 	\caption{\label{IPR-nomarket}(Color on line) The inverse participate ratio (IPR) as a function of eigenvalues with out the largest eigenvalue for different multifractal order $q$ and detrending scale $s$. The red line is the inverse participate ratio 
 	for random matrix with value $\langle I_{k}\rangle=3/N$.}
 \end{figure}
 
\begin{figure}
	\includegraphics[width=\textwidth]{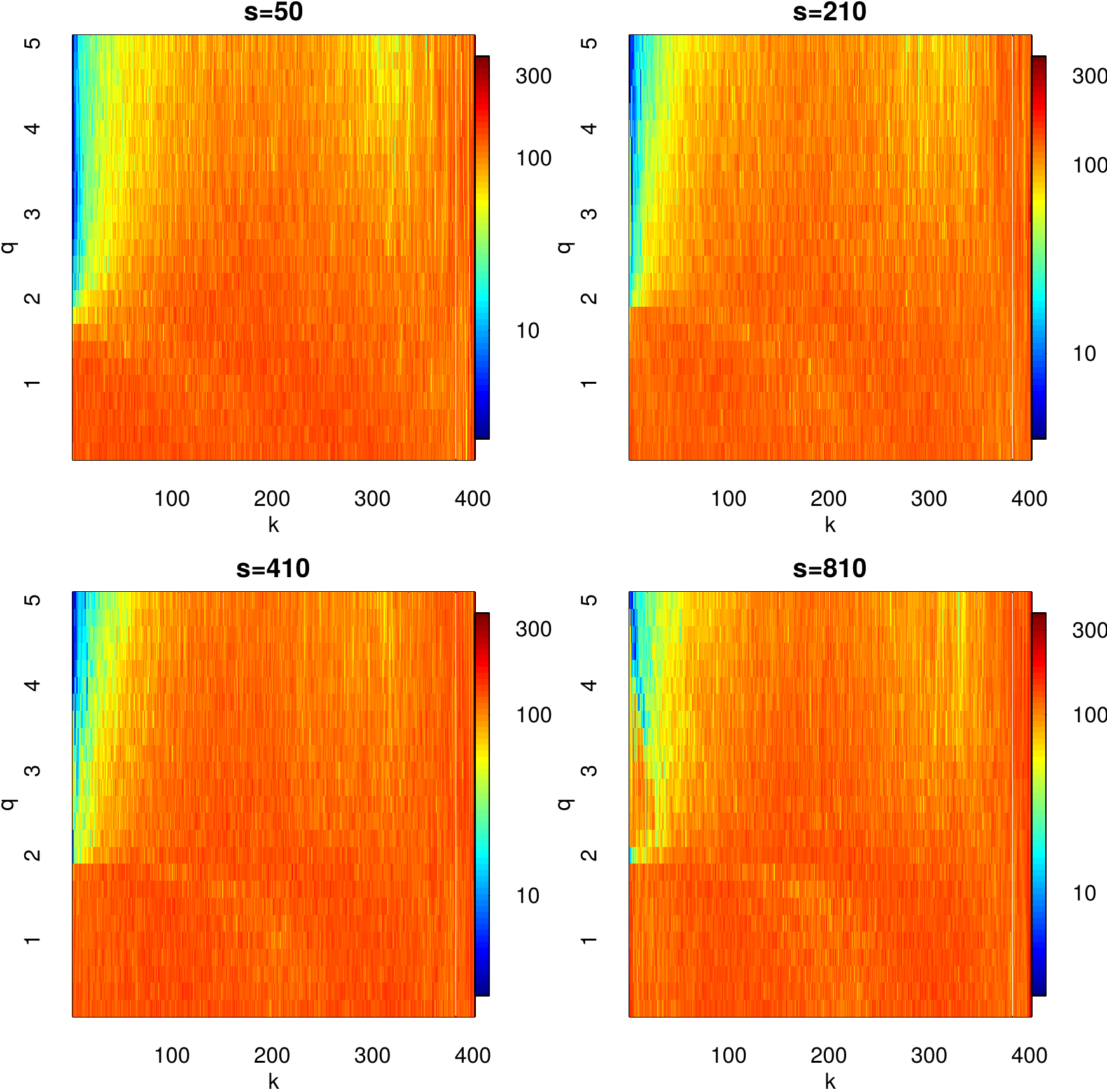}
	\caption{\label{PR}(Color on line) The participate ratio $1/I_{k}$ as a function of $q$ and $k$.$k$ is the label of the eigenvalues $\lambda_{k}$. Here we set detrending scale $s= 50, 210, 410, 810$.}
\end{figure}

As explained above, the inverse participate ratio
quantifies the reciprocal of the number of eigenvector components that contribute significantly.
Here we give the inverse participate ratio of the q-dependent cross-correlation matrices 
at different multifractal orders and detrending scales in Fig.\ref{IPR-nomarket}. We 
present the inverse participate ratio without the largest eigenvalue for better 
visualization. Note that there is a transition in the IPR $I_k$ for small and large multifractal order $q$. When $q\leq 2$, the small eigenvalues are dominant by relatively 
small proportion of stocks with larger IPR. It can be validated using the participate ratio $1/I_k$ in Fig. \ref{PR}, which is the participate ratio for those 
small eigenvalues less than 50. For medium and large eigenvalues
the participate ratio are larger than 200.
\begin{figure}[ht]
		\includegraphics[width=\textwidth]{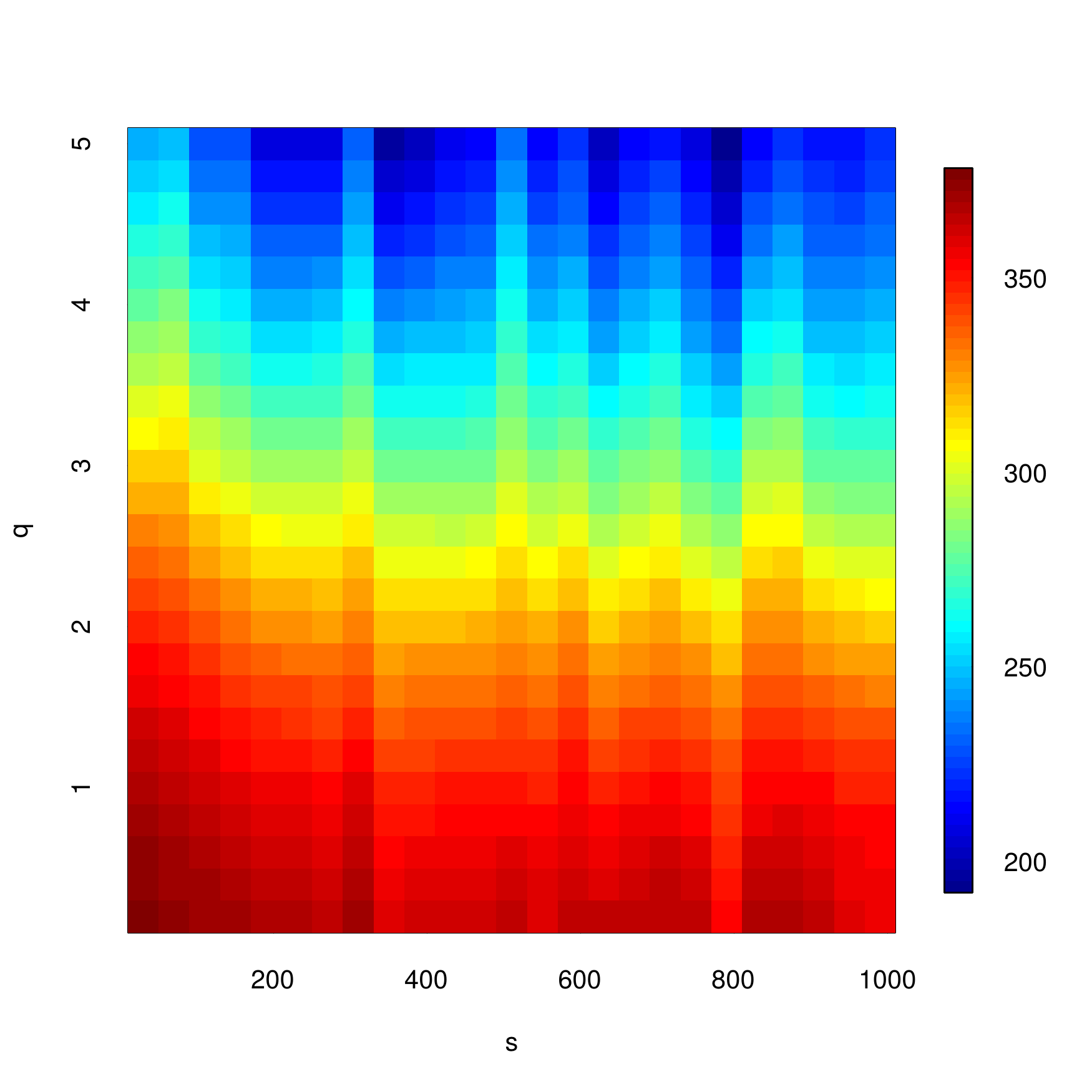}
		\caption{\label{PR1}(Color on line) The participate ratio(PR) $1/I_{k}$ of the largest eigenvalue $\lambda_{401}$ for different multifractal order $q$ and detrending scale $s$.}
\end{figure}

Fig.\ref{PR1} shows the participate ratio
$1/I_{k}$ for the largest eigenvalue. The largest participate ratio for $q<2$ is 376 which approaches the system size $N=401$. When $q\geq 2$, the participate ratio 
for largest eigenvalue decrease rapidly and has a value of 200. The striking difference in the contribution
number of the largest eigenvalues for different fluctuations implies that the 
collective behavior of small fluctuations ($q<2$) are more homogeneous 
(large participate ratio).
Fig.\ref{PR} shows the heat map of the participate ratio $1/I_{k}$ at different multifractal orders
$q$ when $s = 50, 210, 410, 810$. $k$ is the label of the eigenvalue $\lambda_{k}$. When 
$q\geq 2$, the 
participate ratio for small eigenvalues(small $k$) are very small suggesting that the small
eigenvalues contain useful information. Only a very small set of stocks contributed to the smallest eigenvalue. We can verify this using the eigenvector component contribution 
in Fig. \ref{eigenComp1-2}. When $q\geq 2$ the small eigenvalues are dominated 
by a few sectors. This has implications relevant to portfolio optimization. In general, the pattern of collective behavior for small fluctuations differs from that of large fluctuations.

\subsection{PMFG analysis}
The planar maximally filtered graph (PMFG) has been used to analyze the structure and dynamics
of stock market in times of crisis\cite{Song2011,Zhao2015b}, and it effectively captures the sector structures. Here we construct the PMFG networks using q-dependent cross-correlation matrices.

\begin{figure}
	\includegraphics[width=\textwidth]{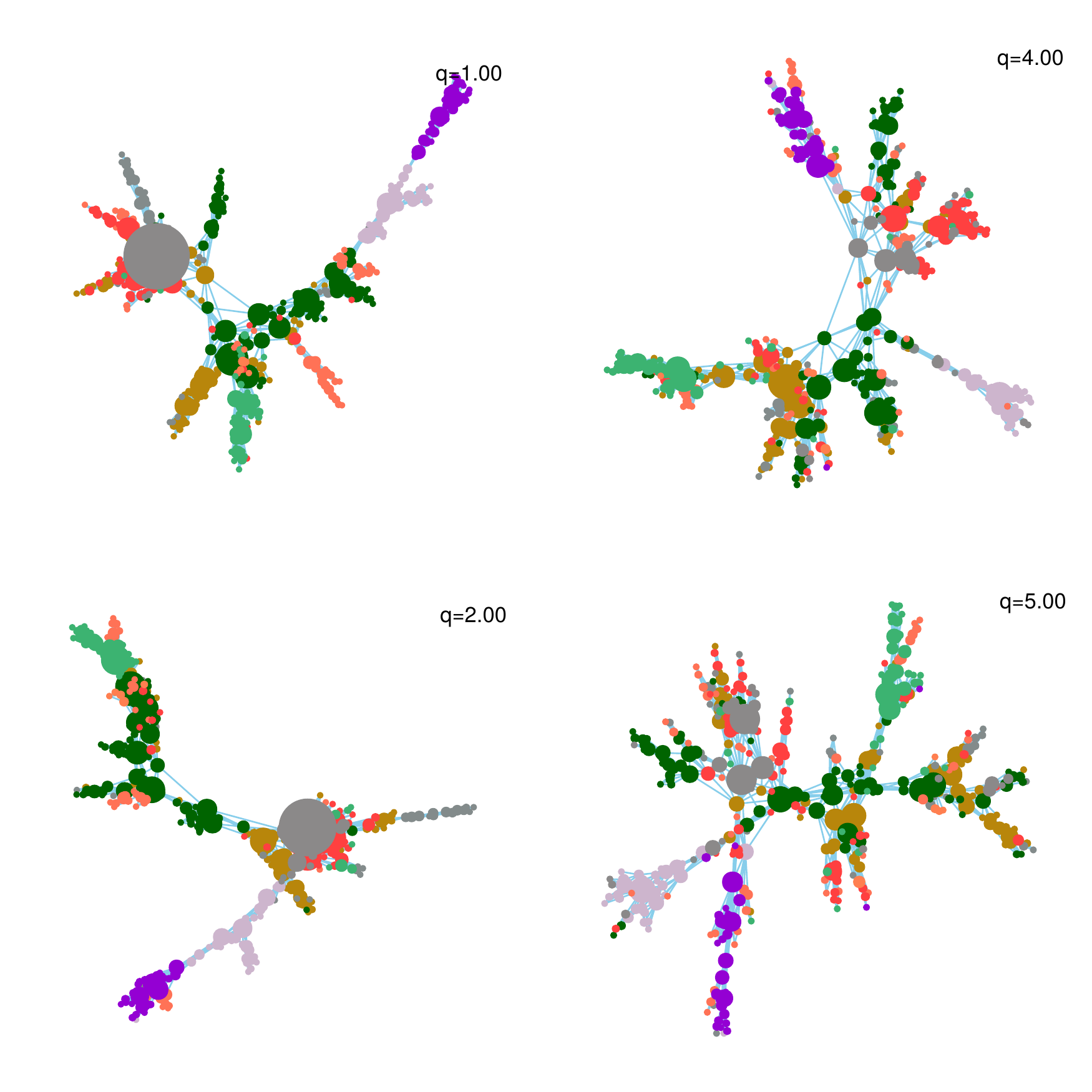}
	\caption{\label{network}(Color on line) The PMFG networks for different orders $q$. We set detrending scale $s=110$ here. Different vertex colors represent different sectors. The vertex size is proportion to the degree of each vertex.}
\end{figure}
Fig \ref{network} shows the networks constructed using PMFG algorithm.
The sector structure for small $q$ is clearer than those for large $q$. Recently Kawpen et al.\cite{Kwapien2016} constructed the minimum spanning trees using the q-dependent cross-correlation matrices. Some hidden structures were found using minute datasets. Here we find that when $q\leq 2$, a hub stock emerges, but when $q>2$, the degree heterogeneity  becomes weak. 
Especially, when $q\leq 2$ the dark green nodes (the stocks form Financial sector) are very close with each other. However, when $q>2$, the links between financial 
sector stocks loosen. Those characteristics qualitatively agree with the results from \cite{Kwapien2016} in which they discover a star like minimum spanning tree structure when $q\leq 2$.

\begin{figure}
	\includegraphics[width=\textwidth]{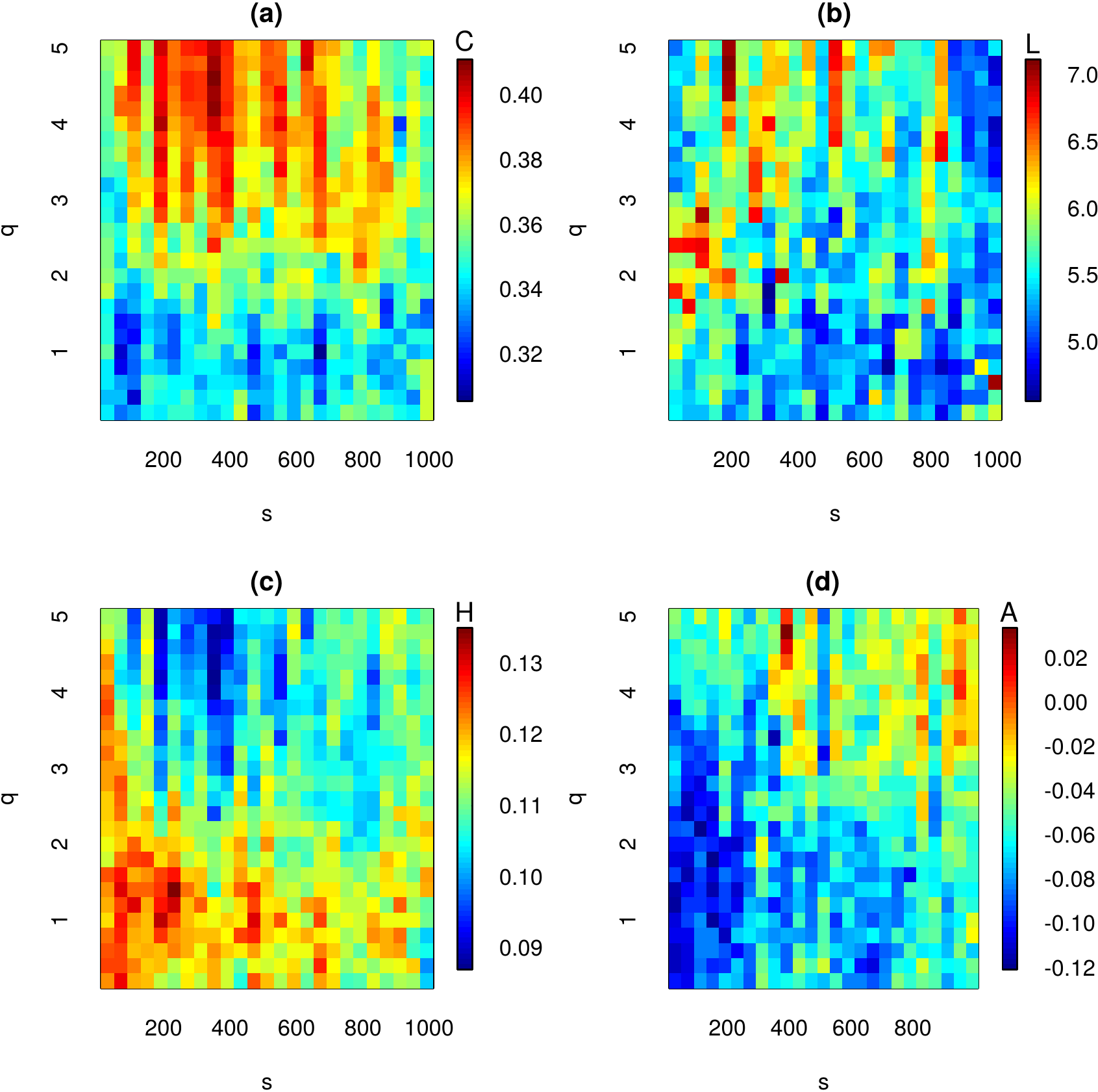}
	\caption{\label{topology}(Color on line) The topological quantities of the PMFGs at different order $q$ and scale $s$. (a) is the clustering coefficient $C$, (b) is the shortest path length $L$, (c) is the heterogeneity index $H$, (d) is the assortativity $A$.}
\end{figure}
To quantify the influence of 
the fluctuations on the PMFGs at different multifractal orders $q$ and detrending scales $s$, we calculate the topological
quantities of the PMFGs.
The topological quantities of the PMFGs are presented in Fig\ref{topology}. Fig\ref{topology} (a) shows that the clustering coefficient $C$ of
PMFGs increases as the multifractal order $q$ increases.
The clustering coefficient is large when the
detrending scale is short. The shortest path length $L$ has been shown in Fig\ref{topology}(b).
The shortest path length is large for large $q$ and short $s$. 
Fig\ref{topology} (c) is the heterogeneity
index $H$\cite{estrada2010quantifying}, which quantifies the heterogeneity level of the PMFGs. It is analogous to the power law index of the scale-free network. It is known that the heterogeneity of BA 
network is 0.11. We notice that for small $q$ the heterogeneity of the PMFG network is larger 
than BA network. This means the structure of the PMFG networks for small multifractal orders 
$q$ are extremely heterogeneous. we also show the assortativity $A$ of the PMFGs at 
Fig\ref{topology} (d). The negative assortativity for $q<2$ gives a hint about the dis-assortative 
structure in which hub stocks tend to connect with the small
degree stocks. When $q>2$ the assortativity approaches 0. This indicates that for large $q$
the connections are more evenly distributed(see Fig. \ref{network}).
In a network with $q>2$ the degree of the hub stocks are smaller than those hubs in 
networks with $q<2$. From the variation of topological quantities, we can infer that
for small fluctuations (small $q$) at short time scale (small $s$), there exist some 
leading stocks. But for large fluctuations (large $q$) and long time scale (large $s$)
, stocks are correlated uniformly.
To sum up, from those topological quantities a obvious structure change is evident 
which gives an indication about the collective behaviors difference among 
fluctuations of different magnitudes at varying time scales.

\section{Application\label{sec4}}
We are now exploring the possibility of using the q-dependent PMFG networks to 
improve the performance of portfolio optimization under the Markowitz portfolio 
framework\cite{Markowitz1952}. First we briefly introduce the Markowitz 
portfolio theory and then we use some centrality metric to choose portfolio from the 
PMFG networks. Considering a portfolio $\Pi(t)$ of stocks with return $r_i,i=1\ldots 
m$, $m$ is the portfolio size, i.e., the number of stocks in the portfolio. The 
return on $\Pi(t)$ of stocks is
\begin{align*}
\Pi(t)=\sum\limits_{i=1}^{m}\omega_{i}r_{i}(t),
\end{align*}
where $\omega_{i}$ is the fraction of wealth invested in stock $i$. The fractions $\omega_{i}$ are normalized such that $\sum\limits_{i=1}^{m}\omega_{i}=1$. The risk in holding the portfolio $\Pi(t)$ can be quantified by the variance \\
\begin{align*}
	\Omega^{2}=\sum\limits_{i=1}^{m}\sum\limits_{j=1}^{m}\omega_{i}\omega_{j}C_{ij}\sigma_i\sigma_j,
\end{align*}
here $C_{ij}$ is the Pearson cross-correlation between $r_i$ and $r_j$, and $\sigma_i$ and $\sigma_j$
are the standard deviations of $r_i$ and $r_j$. To find an optimal portfolio, we maximize the return of the portfolio $\Phi=\sum\limits_{t=1}^{T} \Pi(t)$ under the constraint
 that the risk on the portfolio is some fixed value $\Omega^{2}$. Maximizing $\Phi$ 
 subject to these two constraints which is equivalent to a quadratic 
 optimization problem
\begin{align*}
	\omega^{T}\Sigma\omega-q*\mathrm{R}^{T}\omega 
\end{align*}
Here $\Sigma$ is the covariance matrix of the return matrix $\mathrm{R}$ mentioned in 
the previous context (now with dimension $L\times m$). The parameter $q$ is the risk tolerance $q\in [0,\infty)$. If we 
set large $q$ we have strong tolerance to the risk, which leads to large expected return. 
The optimal portfolios can be represented as a plot of the return $\Phi$ as a function 
of risk $\Omega^2$ which is known as the efficient frontier. Here we do not use the 
q-dependent cross-correlation coefficient in the risk metric $\Omega^2$. We only use the 
q-dependent PMFG networks to select $m$ stocks and then the traditional Markowitz
portfolio theory is used to quantify the performance of the portfolio.
It has shown that portfolio selected from the PMFG networks constructed from Pearson cross-correlation matrix using some centrality measures perform very well\cite{Pozzi2013}. Here we first calculate the centrality scores defined by\\
\begin{align*}
	\eta=\frac{C_{D}^{w}+C_{D}^{u}+C_{BC}^{w}+C_{BC}^{u}-4}{4\times (N-1)}+\frac{C_{E}^{w}+C_{E}^{u}+C_{C}^{w}+C_{C}^{u}+C_{EC}^{w}+C_{EC}^{u}-6}{6\times (N-1)}
\end{align*}
where $C_{D}^{w}$ is the ranking of weighted Degree (D) and $C_{D}^{u}$ is its 
unweighted counterpart. The other centrality metrics are Betweenness Centrality ($BC$), Eccentricity ($E$), Closeness ($C$), 
Eigenvector Centrality ($EC$). A portfolio construct using the central
(peripheral) stocks are those with very high (low) centrality value $\eta$. 
A complete description of this composed centrality metric is provided in
Ref.\cite{Pozzi2013}. Actually, the choice of the centrality metric does not significantly effect the final results.
\begin{figure}
	\includegraphics[width=\textwidth]{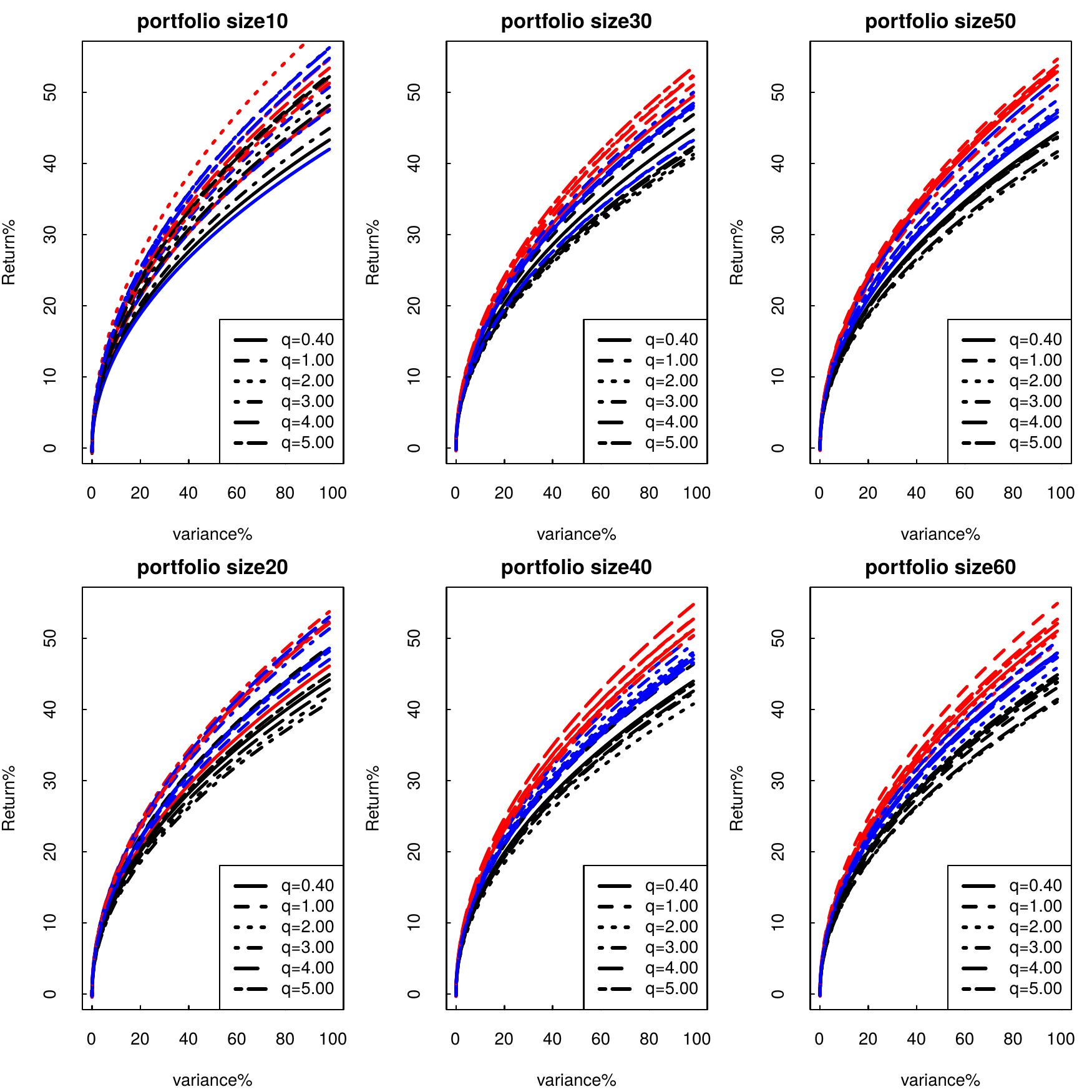}
	\caption{\label{eff}(Color on line) The efficient frontier for different portfolio size $m=10, 20, 30, 40, 50, 60$. The red, blue and black lines are efficient frontiers for those portfolio constructed using peripheral, random selected and central stocks in the PMFG networks.}
\end{figure}
Here in Fig.\ref{eff}, we show the efficient frontiers calculated from those portfolio 
constructed using central (black lines), peripheral (red lines) and random (blue lines)
stocks with different multifractal orders $q$. We have tested on the portfolio size $m=10,20,30,40,50,60$ and calculated the average return value for all the detrending scales $s$ at one specific risk value to show the effect of fluctuations at different magnitudes. It is very
clear that for different portfolio size from 10 to 60 stocks, the peripheral 
portfolio (red lines) are always the best performed. The performance of the central portfolio is even worse than the random portfolio.

\begin{figure}
	\includegraphics[width=\textwidth]{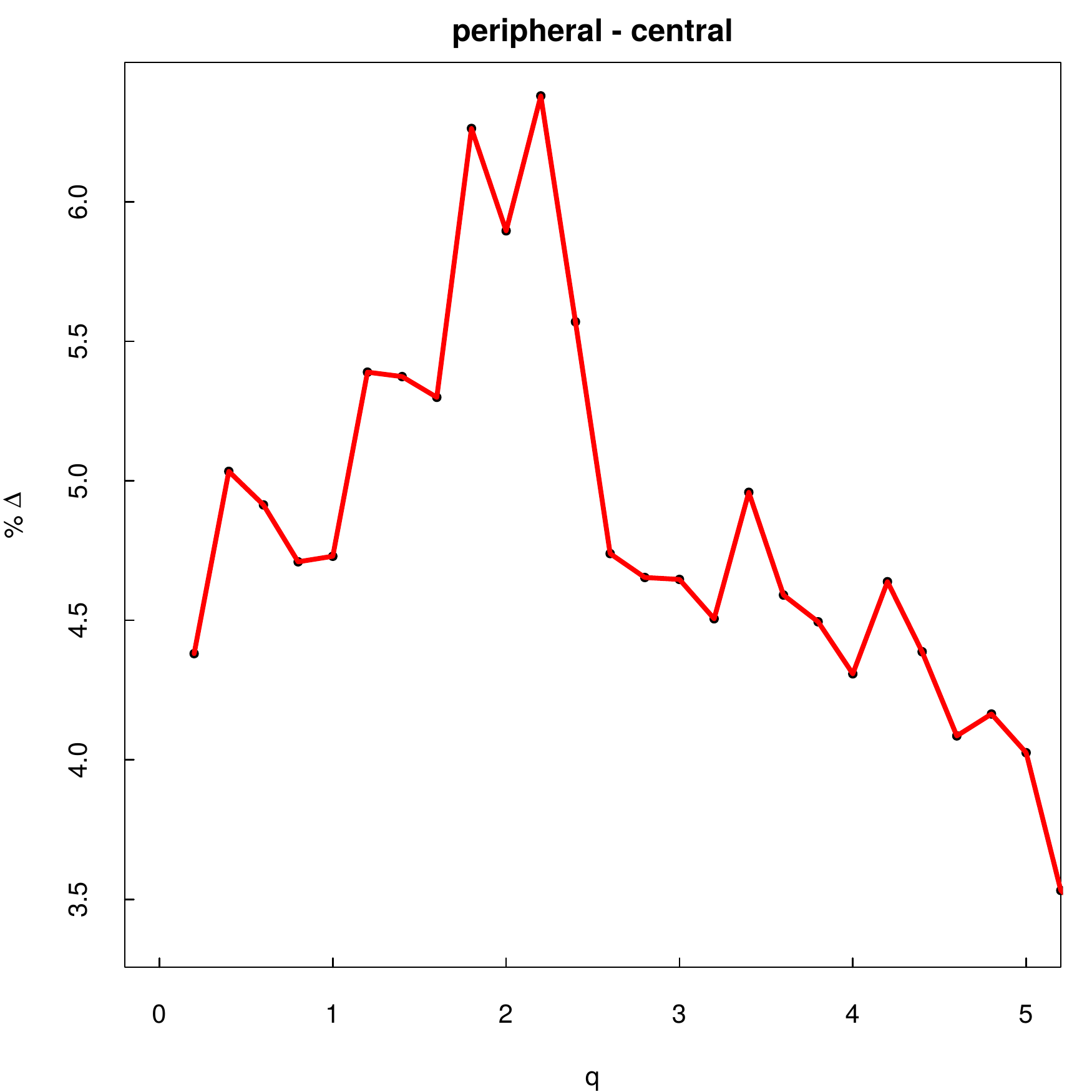}
	\caption{\label{diff}(Color on line) The difference between the return of peripheral and the return of central portfolio as a function of multifractal order $q$. }
\end{figure}
We then calculate the return difference between peripheral and central portfolio 
$\Delta = 	\Phi_{p} - \Phi_{c}$ ($p$ and $c$ are peripheral and central portfolios, 
respectively) as a function of the multifractal order $q$ in Fig.\ref{diff}. Here we 
use multifractal orders from 0.2 to 10 to identify the optimal $q$. It's very obvious
that the peripheral portfolio outperforms the central portfolio most around 
multifractal order $q=2$ and exhibits a greater than $\% 7$ superiority. This may gives a 
hint that we should trade based on moderate fluctuation with higher return
and lower risk. The results above indicate the 
potential of utilizing the q-dependent cross-correlation matrix as a new portfolio
optimization tool.

\section{Conclusion\label{sec6}}
In this paper, we have employed the q-dependent cross-correlation coefficient to analyze 
the cross-correlation among fluctuations at different magnitudes for stock market. With the help of random matrix theory and
complex network theory, we analyze the cross-correlation matrices of the stock market for different magnitude of fluctuations. We find that the cross-correlation among small
fluctuations are stronger than large ones. There are more deviating eigenvalues  
for large fluctuations than that for small fluctuations. Analyzing the inverse 
participate ratio and the eigenvector contribution, we find that the small eigenvalues of the cross-correlation matrices for large fluctuations are dominated by
a small number of industry groups. This is similar to those large deviating eigenvalues that are also dominated by a small number of industry groups. Thus we conclude that small eigenvalues of the q-dependent cross-correlation matrices also carry some genuine information, which seems very counterintuitive. The 
complex network representation have also validate the correlation difference between
small and large fluctuations. The network structure are more heterogeneous and 
dis-assortativity for the network constructed from small fluctuations which means the 
existence of leading stocks for small fluctuations. We then utilize
the network centrality as a portfolio selection metric. Under the Markowitz portfolio 
theory, we find that the portfolio of the peripheral stocks always outperforms the
portfolio of central stocks. Optimal multifractal order $q$ with the largest return
difference approaching $\% 7$ is approximately $q=2$. This may be used as a 
new portfolio optimization tool. Thus our investigation about the 
cross-correlation among stocks with different magnitude of fluctuations have demonstrated the 
huge difference between large and small fluctuations of stock market. They are 
regulated by different non-linear correlation structures. Those results expands our 
understanding of the collective behavior of the stock market.
\section{Acknowledgments}
This work is supported in part by the Programme of Introducing Talents of Discipline
to Universities under grant NO. B08033 and and the program of China Scholarship Council (No. 201606770023).
\bibliographystyle{unsrt}

 \end{document}